\documentclass[12pt]{article}
\usepackage{graphicx,epsfig}
\usepackage{amsfonts}
\usepackage{theorem}
\usepackage{amsmath}
\usepackage{amssymb}
\usepackage{a4}
\def\al{\alpha}
\def\th{\theta}
\begin{document}
\title{The flavor physics in unified gauge theory
 from an $S_3\times{\cal P}$ discrete symmetry}
\author{Stefano Morisi and Marco Picariello
\\\small INFN - Milano and Universit\`a degli Studi di Milano,
\\\small Dipartimento di Fisica - Sezione Teorica - via Celoria 16 - 20133 Milano - Italy}
\maketitle
\abstract
We investigate the phenomenological implication of the discrete
symmetry $S_3\times{\cal P}$ on flavor physics in $SO(10)$ unified theory. 
We construct a minimal renormalizable model which
reproduce all the masses and mixing angle of both quarks and leptons.
As usually the $SO(10)$ symmetry gives up to relations between the down
sector and the charged lepton masses.
The underlining discrete symmetry gives a contribution (from the charged
lepton sector) to the PMNS mixing matrix which is bimaximal.
This gives a strong correlation between the down quark and charged lepton
 masses, and the lepton mixing angles.
We obtain that the small entries
$V_{ub}$, $V_{cb}$, $V_{td}$, and $V_{ts}$ in the CKM matrix
are related to the small value of the ratio
$\delta m^2_{sol}/\delta m^2_{atm}$:
they come from both the $S_3\times{\cal P}$
 structure of our model and the small ratio of the other quark
 masses with respect to $m_t$.
%
%
\section{Introduction}
It is well know that there could be some theoretical relations between
quark and lepton masses, however apparently Nature indicates that
the lepton mixing angle should be completely uncorrelated to the quark mixing
angles.
Recent neutrinos experimental data show that in first approximation,
the lepton mixing PMNS matrix is  tri-bimaximal, i.e. the
atmospheric mixing angle is maximal, $\theta_{13}\approx 0$ and the solar
angle is $\theta_{12}\approx \arcsin( 1/\sqrt{3})$.
The tri-bimaximal matrix
follow in natural fashion in models invariant under discrete
symmetry like $S_3$ which is the permutation symmetry of  tree
object~\cite{Caravaglios:2005gw}.
These motivations suggest us to consider discrete symmetries
in extensions of the unified version of the SM.
In literature are investigated both unified models based on extension
of the standard model such us $SO(10)$~\cite{Georgi:1974sy}
symmetry with~\cite{Barbieri:1996ww} or without~\cite{Dutta:2004hp}
continuous flavor symmetries, and not unified models based on discrete
symmetries~\cite{Altarelli:2005yp}.
Although some of them appear to be promising in understanding the flavor
physics and unification~\cite{Chen:2005jt}
we are still far from an unitarity vision of the flavor
problem~\cite{Caravaglios:2002br}.
Because the $S_3$ flavor permutation symmetry is hardly broken in the
phenomenology, in this paper we study a model invariant under the
$SO(10)\times S_2\times{\cal P}$ group,
where the $S_2\times{\cal P}$ group is the discrete flavor symmetry.
We analyse the phenomenological implication of such discrete
symmetry on flavor physics and our aim is to
construct a minimal renormalizable model which reproduce all the
masses and mixing angles of both quarks and leptons.
The $S_2\times{\cal P}$
 symmetry implies that the resulting mass matrices of the
fermion are not general, but depending each one on 5 free parameters only.
Together with the assumption that the two Higgs in {\bf 10} couple to fermions
with a Yukawa matrix of rank one~\cite{Altarelli:2005yp},
we obtain that the left mixing
matrices are all bimaximal with the remaining mixing angle small.
This implies that the CKM is almost diagonal in the $S_2$ exact case.
In our model the tri-bimaximal PMNS mixing matrix is
achieved by rotating the low energy neutrino mass matrix.
In a very surprising way we obtain that the small entries
$V_{ub}$, $V_{cb}$, $V_{td}$, and $V_{ts}$  are related
 to the small value of the ratio
$\delta m^2_{sol}/\delta m^2_{atm}$ (coming from both
 the $S_2\times{\cal P}$
 structure of our model and the small ratio of the quark
 masses with respect to $m_t$).
On the other side when the $S_2$ symmetry is dynamically broken 
only the Cabibbo angle becomes relevant.
\section{Our model}
In $SO(10)$ all the fermion fields, with the inclusion of the
right-handed neutrino, can be assigned to the ${\bf 16}$
dimensional multiplet.
We introduce the three possibilities to construct renormalizable invariant
mass terms
\begin{equation}
{\bf 16~16~10},~~~~{\bf16~16~120},~~~~{\bf 16~16~\overline{126}}\label{3}
\end{equation}
where ${\bf 10,~120,~\overline{126}}$ are Higgs scalar fields.
We consider the patter breaking of $SO(10)$ into the Standard
Model through the Pati-Salam $G_{224}$ group.
From the branching rules of $SO(10)~\supset~G_{224}$,
 it can be show that the non negligible Majorana
mass term  can arise only from the third interaction in (\ref{3})
with the  ${\bf \overline{126}}$ scalar field.

We introduce a ${\bf 16}^i$ multiplet for each flavor $i$.
We split the fermions into the $\{1,2\}$, which are taken doublet under
$S_2$, and $\{3\}$, $S_2$ singlet.
We add two Higgs scalars  ${\bf \overline{126}}^\al$,
and a ${\bf 120}$.
We assume that the two fields ${\bf \overline{126}}^\al$
 form a doublet under $S_2$, and
we write the $SO(10)\times S_2$ invariant Lagrangian
\begin{eqnarray}   \label{eq:lag}
L_{\text{yuk}}^{b}&=&
{\bf I}_{ij}~{\bf 16}^{i}~{\bf 16}^{j}~{\bf 10}
+ {\bf g}_{ija}{\bf 16}^{i}~{\bf 16}^{j}~{\bf \overline{126}}^\al
\\&&
+{\bf A}_{ij}{\bf 16}^{i}~{\bf 16}^{j}~{\bf 120}
+{\text{h.c.}}
\nonumber
\end{eqnarray}
The flavor indices $\{i,j\}$ run over $\{1,2,3\}$, and the $\al$
 over $\{1,2\}$.
We introduce a parity operator ${\cal P}$ under which the fields
 transform as follow:
\begin{eqnarray*}
{\cal P} {\bf 16}^\al  = - {\bf 16}^\al
&&
{\cal P} {\bf 16}^3    =  {\bf 16}^3\\
{\cal P} {\bf 126}^\al = {\bf 126}^\al
&&
{\cal P} {\bf 120}     = - {\bf 120}\\
{\cal P} {\bf 10}      = {\bf 10}
\end{eqnarray*}
The symmetric tensor ${\bf g}_{ija}$, and the antisymmetric matrix ${\bf A}$
are the most general $S_2\times {\cal P}$ invariant and are given by
\begin{eqnarray}
\!{\bf g}_{ij1}=\left(
\begin{tabular}{ccc}
b & d & 0 \\
d & e & 0 \\
0 & 0 & f
\end{tabular}
\right)
\!,
{\bf g}_{ij2}=\left(
\begin{tabular}{ccc}
e & d & 0 \\
d & b & 0 \\
0 & 0 & f
\end{tabular}
\right)
\!,&&\!
{\bf A}=A\left(
\begin{tabular}{ccc}
0       & 0     & -1\\
0       & 0     & -1\\
1       & 1     & 0\\
\end{tabular}
\right)
\nonumber
\end{eqnarray}
while, as it will be clarified in the next section, ${\bf I}$ will not be
taken the most general symmetric matrix invariant under our flavor group.
The coupling constants in ${\bf g}$, ${\bf A}$, and ${\bf I}$
are assumed to be small enough to avoid problem with respect
the electroweak precision tests. They are all of the same order
of magnitude.

The decomposition of the ${\bf 10}$, ${\bf 120}$, and ${\bf \overline{126}}$
representations under the group $SU_L(2)\times SU_R(2)\times SU_c(4)$ are
\begin{eqnarray}
{\bf 10} &=&(2,2,1)+(1,1,6)
\nonumber\\
{\bf 120}&=&(2,2,1)+(1,1,\overline{10})+(1,1,10)+(2,2,15)+
\nonumber\\&&\quad (1,3,\overline{6})+(3,1,6)
\nonumber\\
{\bf \overline{126}}&=&(\overline{3},1,\overline{10})+(1,\overline{3},10)+(2,2,15)+(1,1,6)
\nonumber
\end{eqnarray}
Under the same group the ${\bf 16}$ decompose in
$(2,1,4)_L$ and $(1,2,\overline{4})_R$. Then the Dirac mass terms
decompose as follow
\begin{equation}
(2,1,4)_L \times (1,2,\overline{4})_R = (2,2,1)+(2,2,15)\label{4}
\end{equation}
and the Majorana mass terms are
\begin{equation}
(1,2,\overline{4})_R \times (1,2,\overline{4})_R = (1,3,\overline{10})+
(1,1,\overline{10})\label{5}
\end{equation}
where in (\ref{5}) we have neglected the terms containing the {\bf 6}
of SU$(4)$ which break the color symmetry. The Majorana mass cames
from the $(1,\overline{3},10)$ component of ${\bf \overline{126}}$
and the Dirac mass cames from the (2,2,1) and (2,2,15)
components respectively of the ${\bf 10}$, ${\bf 120}$ and ${\bf
\overline{126}}$.
We assume, by using the experimental constrains coming out from
the electroweak precision tests of the Standard Model, that there are  
only two light Higgs doublets. In the mass bases for the Higgs,
two of the {\em vev}s are assumed to be
$\approx$ 100 GeV ($k^u$ and $k^d$)
and all the others {\em vev}s are much smaller the 100 GeV.

We are able now to write down the mass matrices of the quarks
and leptons that follow from the model given by the
Yukawa interactions (\ref{eq:lag})
\begin{subequations}\label{eq:masses}
\begin{eqnarray}
M^u&=&k^u~{\bf I}+ {\bf \Delta^u}+(q^u_s+q^u_{adj}){\bf A}
\label{mu}\\
M^d&=&k^d~{\bf I}+ {\bf \Delta^d}+(q^d_s+q^d_{adj}){\bf A}
\label{md}\\
M^l&=&k^d~{\bf I}- 3~ {\bf \Delta^d}+(q^d_s-3q^d_{adj}){\bf A}
\label{ml}\\
M^{\nu}&=&k^u~{\bf I}-3~ {\bf \Delta^u}+(q^u_s-3q^u_{adj}){\bf A}
\label{mnu}\\
M^{\nu_R}&=&{\bf \Phi}\label{mr}
\end{eqnarray}
\end{subequations}
where $k^{u,d}$ are the {\em vev}s of the two standard Higgs doublets
of (2,2,1) in ${\bf 10}$, 
the $q^{u,d}$ are the {\em vev}s
in ${\bf 120}$ and the index {\em s} and {\em adj} stand for $SU_c(4)$ 
singlet and adjoint representation \cite{Dutta:2004hp}.

The matrices ${\bf \Delta}^{u,d}$, and ${\bf \Phi}$ are 
\begin{eqnarray}
{\bf \Delta^{u,d}}=\left(
\begin{tabular}{ccc}
b$\delta_1$+e$\delta_2$& d$(\delta_1+\delta_2)$ & 0\\
d$(\delta_1+\delta_2)$ & e$\delta_1$+b$\delta_2$& 0\\
0                      & 0                      & f$(\delta_1+\delta_2)$
\end{tabular}
\right)^{u,d}
\nonumber\\
{\bf \Phi}=\left(
\begin{tabular}{ccc}
b$\phi_1$+e$\phi_2$& d$(\phi_1+\phi_2)$ & 0\\
d$(\phi_1+\phi_2)$ & e$\phi_1$+b$\phi_2$& 0\\
 0                 & 0                  & f$(\phi_1+\phi_2)$
\end{tabular}
\right)
\nonumber
\end{eqnarray}
where $\delta_\al^{u,d}$ are the {\em vev}s of the (2,2,15),
and $\phi_\al$, are the {\em vev}s of $(1,\overline{3},10)$
component in the two ${\bf \overline{126}}^\al$s.
In the case that $\delta_1^{u,d}=\delta_2^{u,d}$, and $\phi_1=\phi_2$ than
we obtain that the $S_2$ discrete symmetry is unbroken. However, as we will
show in the following sections, this is not the choice taken by Nature.
For example this case will give a wrong Cabibbo mixing angle.
To obtain a good masses and mixing angles pattern we must require
that $S_2$ is dynamically broken.
\section{Our ansatz}
Up to now, the only assumption we did is the fact that there is a
factor 100 between the two kind of {\em vev}s. This allows us to fit
the big top mass.

By studying our model we find that we have more freedom than what we need
to reconstruct all the masses and mixing angles in quark and lepton sector.
For this reason we assume that the ${\bf I}$ matrix is
not the most general one under the $S_2\times{\cal P}$ symmetry.

In fact, although the most general $S_2\times {\cal P}$
 invariant symmetric matrix is of the form
$$
\left(
\begin{tabular}{ccc}
b & d & 0 \\
d & b & 0 \\
0 & 0 & f
\end{tabular}
\right)\,.
$$
We make the ansatz that the matrix ${\bf I}$ is given by
$$
{\bf I}\propto \left(
\begin{tabular}{ccc}
0               & 0     & 0\\
0               & 0     & 0\\
0               & 0     & 1\\
\end{tabular}
\right)\,.
$$
The reason for this ansatz is related to the high value of the top mass.
The ${\bf I}$ gives (under the assumption that the $k$'s are
much bigger than all the other {\em vev}s) the top, bottom, and tau masses,
and the hierarchy between these and the other masses is given 
by the $\Sigma/k$ and $q/k$ ratios.
Maybe it is possible to justify our ansatz 
from a symmetry bigger than $S_2$ which constrains
the matrix ${\bf I}$ 
(such as a modification of the $U(2)$ in~\cite{Barbieri:1996ww})
but we will not investigate this point in this paper.

For simplicity, we rewrite the $\Delta$s (and equivalently the $\Phi$)
matrices as 
$$
\left(
\begin{matrix}
 \Sigma_1 & \Sigma   & 0 \cr
 \Sigma   & \Sigma_2 & 0 \cr
   0      &    0     & \Sigma_3 \cr
\end{matrix}
\right)
$$
Notice that the $S_2$ symmetry implies $\delta_1=\delta_2$ and then that 
$\Sigma_1=\Sigma_2$.
Moreover the entry $\{3,3\}$ is irrelevant
(except that in $M^{\nu_R}$), because 
the presence of the $k$'s in the mass matrices in eqs.~\ref{eq:masses}.
\section{Charged leptons and down quarks masses}
We know that at the unification scale the relation between
the quark and lepton masses are \cite{Georgi:1979ga}
\begin{subequations} \label{datam}
\begin{eqnarray}
m_\tau &\approx& m_b\,,\label{datamb}\\
m_\mu&\approx& 3m_s\,\label{datams}\\
m_e&\approx&\frac{1}{3}m_d\label{datamd}
\end{eqnarray}
\end{subequations}
It is easy to see that, due to our structure of the mass matrices, we obtain
automatically the relation~(\ref{datams}).

From the equations (\ref{md}) and (\ref{ml}) we obtain the relation
\begin{equation}
3~M^d+M^l~=~4~k^d~{\bf I} + 4~q^d_s~{\bf A}\,.\label{mdml1}
\end{equation}
If the {\bf 120} do not couple to the fermions eq. (\ref{mdml1})
gives wrong relation between lepton and quark masses.
This is the reason of the introduction of the {\bf 120} Higgs
fields in the Lagrangian (\ref{eq:lag}). 
While we need the $SU(4)$ singlet of the {\bf 120} to obtain good relations
between the charged lepton and down quark masses, in the follow we
will assume that the {\em vev} of the $SU(4)$ adjoint into the {\bf 120} is
negligible and we will omit it.

From the fact that all the other {\em vev}s are much smaller that $k^d$'s,
and by assuming that for the moment $\Sigma_2^d=\Sigma_1^d$,
we get that the eigenvalues of
\begin{eqnarray*}
{\bf M^{d}}=
\left(
\begin{matrix}
 \Sigma^d +\Sigma^d_2 & \Sigma^d              & -q^d \cr
 \Sigma^d             & \Sigma^d+\Sigma^d_2   & -q^d \cr
q^d                   & q^d                   & k^d +\Sigma^d_3 \cr
\end{matrix}
\right)
\end{eqnarray*}
 are approximately
$$
\{m_d,m_s,m_b\}=
\left\{
\frac{\Delta^d}{k^d}\,,\quad \Sigma^d_2\,,\quad k^d
\right\}\,.
$$
where $\Delta^d$ is a function of the {\em vev}s given by
\begin{equation}\label{eq:delta}
\Delta^d =  2 \left((\Sigma^{d}_2)^2-(q^d)^2\right)
  +\frac{1}{4}(\Sigma^{d}-\Sigma^{d}_3+\Sigma^{d}_2)^2
 + (2 \Sigma^d + \Sigma^d_2)k^d\,.
\end{equation}
Equivalently the eigenvalues of charged leptons matrix
\begin{eqnarray*}
{\bf M^{l}}=
\left(
\begin{matrix}
 -3 \Sigma^d -3 \Sigma^d_2 & -3\Sigma^d         & - q^d \cr
 -3 \Sigma^d            & -3\Sigma^d-3\Sigma^d_2& - q^d  \cr
 q^d                    & q^d                   & k^d -3\Sigma^d_3 \cr
\end{matrix}
\right)\label{clep}
\end{eqnarray*}
are approximately
$$
\{m_e,m_\mu,m_\tau\}=
\left\{
\frac{\Delta^l}{k^d},\quad
 -3 \Sigma^d_2\,,\quad k^d
\right\}                      \,,$$
where $\Delta^l$ is another function of the {\em vev}s.
It is obvious that the experimental relations (\ref{datam}) can be
easily reproduced in our model.
This fix the value of the $\Sigma_2^d$ (the eigenvalue of $M^l$ which
is three times the eigenvalue of $M^d$) to $m_\mu$ at the unification
scale, and $k^d$ gives the value of $m_\tau$ (by neglecting
$\Sigma_3$, the third eigenvalues of
$M^d$ and $M^l$ are equal). Notice that, in spite the relations
between $\Delta^l$, and $\Delta^d$
(but this point should be better investigate, in fact it
could be an evidence for a more fundamental symmetry of the Standard
Model) needed to reproduce the electron and down masses,
up to now, we fitted six experimental masses by using four {\em vev}s.
\section{Lepton mixing angles and structure of the neutrino mass matrices}
\label{se:mixing}
In general the lepton mixing matrix is $V_{PMNS}=U_{lL}^\dagger U_{\nu
L}$, where $U_{lL}$ and $U_{\nu L}$ enter into the diagonalization of
the charged leptons and neutrino mass matrices.
It is straightforward that if charged leptons mass
matrix has the general $S_2$ invariant structure then the $U^l$
matrix has the form \cite{Caravaglios:2005gw}
\begin{eqnarray}\label{Ue}
\left(
\begin{tabular}{ccc}
$-\frac{1}{\sqrt{2}}$ & $a$   & $b$\\
$ \frac{1}{\sqrt{2}}$ & $a$   & $b$\\
   0                  & $N_a$ & $N_b$
\end{tabular}
\right)
\end{eqnarray}
With a mass matrix
\begin{eqnarray}
\left(
\begin{tabular}{ccc}
67.86 & 57.2  & 65\\
57.2  & 47.06 & 65\\
83.2  & 83.2  & 1560
\end{tabular}
\right)
\end{eqnarray}
we obtain
\begin{eqnarray}
\left(
\begin{tabular}{ccc}
 0.64 & -0.77  & -0.057\\
-0.77 & -0.63  & -0.056\\
0.0072& -0.079 & -0.997
\end{tabular}
\right)
\end{eqnarray}
This means that the charged electron mass matrix is diagonalized by
$$
U_e \approx - U_{23}(\th^e) Diag\{1,1,-1\}U_{13}(-\th^e) U_{12}(2\th^e+Pi/4)
$$
where $\th^e\approx 0.07$.
\\
The neutrino mass matrix, which plays a role for the lepton mixing
angles, is the one which comes out from the see saw mechanism, which
in our model is of type I.
In our model the neutrino mixing matrix is again of the form~\ref{Ue},
but, being with an almost exact $S_3$ symmetry, with a column of all
entries of order $\frac{1}{\sqrt{3}}$ given by the singlet under the
$\{1,2,3\}$ permutation group. Moreover the remaining $S_2$ symmetry
implies a column of type $\frac{1}{\sqrt{2}},0,-\frac{1}{\sqrt{2}}$.
\\
With a mass matrix given by
\begin{eqnarray}
\left(
\begin{tabular}{ccc}
7.5  & 3.45 & 4.05\\
3.45 & 1.5  & 4.05\\
4.05 & 4.05 & 6.75
\end{tabular}
\right)
\end{eqnarray}
we obtain
\begin{eqnarray}
\left(
\begin{tabular}{ccc}
-0.19 & -0.73 & -0.66\\
-0.91 & -0.13 & -0.40\\
-0.38 & -0.67 & -0.63
\end{tabular}
\right)
\end{eqnarray}
This means that the neutrino mass matrix is diagonalized by
$$
U_e \approx - U_{23}(\pi/4-\th^\nu) Diag\{-1,1,1\}U_{13}(-\pi/4) U_{12}(\th^\nu-Pi/2)
$$
where $\th^\nu\approx \arcsin(0.22)$.

We see that we obtain the tri-bimaximal PMNS mixing matrix
\begin{equation}\label{eq:PMNS}
\left(
\begin{array}{ccc}
\sqrt{\frac{2}{3}}   &\frac{1}{\sqrt{3}}  &0 \\
-\frac{1}{{\sqrt{6}}}&\frac{1}{{\sqrt{3}}}&-\frac{1}{{\sqrt{2}}} \\
-\frac{1}{{\sqrt{6}}}&\frac{1}{{\sqrt{3}}}&\frac{1}{{\sqrt{2}}}
\end{array}
\right)
\end{equation}
which fit the experimental data~\cite{exp}.
\section{Up quark masses}
Let us now analyze the up quark mass matrix:
$$
\left(
\begin{matrix}
  \Sigma^u + \Sigma^u_1 &  \Sigma^u              & -q^u \cr
  \Sigma^u              &  \Sigma^u+{\Sigma^u_2} & -q^u  \cr
q^u                     & q^u         & {k^u} + {\Sigma^u_3} \cr
\end{matrix}
\right)
$$                                   
His eigenvalues are approximately
$$
\{m_d,m_c,m_t\}=
\left\{
\frac{\Delta^u}{k^u}\,,\quad \Sigma^u_2\,,\quad
k^u
\right\}\,.
$$
where $\Delta^u$ is a function of the {\em vev}s like~(\ref{eq:delta}).
With $k^u$ we fit the experimental values of the top mass.
By using the remaining freedom for the values of the {\em vev}s 
$\Sigma^u_2$
we fit the experimental values of the charm quark masses.
For the up quark mass we have two cases:
if $q$ is small compared to $k^u$, then 
there is a fine tuning between $\Sigma^{u}$ and $\Sigma^{u}_2$.
If $q$ is bigger there is a fine tuning 
which fix the ratio $\Sigma^u/q^u$.
In our fit will use the first situation, and impose that $q$ is much
smaller then $k$.
\section{Neutrino masses and the CKM matrix}
The low energy neutrino masses, coming from the see-saw between the Dirac
and Majorana neutrino mass matrices, depend directly on the $k^u$,
the three $\Sigma^u_i$, and the four {\em vev}s $\phi$.
The small value of the ratio
$\delta m^2_{sol}/\delta m^2_{atm}$
is approximately equal to $-2(q^u)^2/(k^u)^2$.
This fact is coming from both
 the $S_2$ structure of our model and the small ratio between
the other quark masses and $m_t$.

As we told, if the $S_2\times{\cal P}$
 symmetry is exact than the CKM matrix is not
the right one. 
The $S_2\times{\cal P}$ symmetry in our model implies that the left mixing
matrices are all bimaximal with the remaining mixing angle small.
This implies that the CKM is almost diagonal in the $S_2$ exact case.

We observe that the small entries $V_{ub}$, $V_{cb}$, $V_{td}$, and $V_{ts}$
 in the CKM matrix are related to the small value of the ratio
$\delta m^2_{sol}/\delta m^2_{atm}$.
All of them, in our model, are approximately
proportional to a power of $q^u/k^u$.

In our model, the $S_2$ symmetry is broken only in the neutrino-up
sector to fit
the CKM mixing angles and to not destroy the prediction of a bimaximal
PMNS mixing matrix.
In this case, the Cabibbo angle is the only mixing angle hardly related
 to the $S_2$ breaking.
Moreover this breaking introduce a correction for the other entries of
the CKM which goes into the right direction for obtaining
$V_{ub}<<V_{cb}$, and $V_{td}<<V_{ts}$.
Finally we get the following solution for the CKM matrix
\begin{eqnarray*}
\left(
\begin{matrix}
0.9742 & 0.226 & 0.0036\\
0.225  & 0.9735& 0.039\\
0.012  & 0.038 & 0.9992
\end{matrix}
\right)
\end{eqnarray*}  
which agrees very well with the experimental
values~\cite{Eidelman:2004wy}.

The $S_2$ breaking enters now in the determination of
 the $\theta_{sol}$ too. However we are able to impose that the
low-energy neutrino mass matrix is diagonalized by a rotation into the
$\{1,2\}$ family, by using the freedom in the right-handed sector.
In this way it is possible to fit both the experimental constraints about
 the value of $\delta m^2_{atm}$ and $\delta m^2_{sol}$, and the
 observed PMNS mixing matrix given in eq.~(\ref{eq:PMNS}).
However to explore the full predictivety of our model we need a Monte Carlo
simulation.
\section{Conclusions}
In this paper we analysed a model based on $SO(10)$ gauge symmetry
times and $S_3\times{\cal P}$ discrete flavor symmetry.
The aim of this work was to show that there is a symmetry beyond
the lepton and quark masses despite the fact that the CKM and PMNS
matrix are so different.

By using the most general $S_2\times{\cal P}$
invariant Lagrangian with one ${\bf 10}$,
one ${\bf 120}$, and two ${\bf 126}$ Higgs, we are able to reproduce all
the quark and lepton masses and mixing angles.
Moreover, by making an ansatz which allows us to reduce the number of
free Yukawa coupling we are able to construct a model which predict the
usual unification relations between the down and the charged lepton masses.
Our model agree very well with the recent neutrinos experimental
data, that in first approximation
 give the lepton mixing PMNS matrix tri-bimaximal (i.e. the
atmospheric mixing angle is maximal, $\theta_{13}\approx 0$, and the solar
angle $\theta_{12}\approx \arcsin( 1/\sqrt{3})$).
This tri-bimaximal matrix follow in natural fashion in our model.
The $S_2\times{\cal P}$
 symmetry, together with the assumption that the two Higgs
in {\bf 10} couple to fermions with a Yukawa matrix of rank one,
implies that the left mixing
matrices are all bimaximal with the remaining mixing angle small.
This implies that the CKM is almost diagonal in the $S_2$ exact case.

By giving as input the three charged lepton masses and the
down quark mass, we obtain as output the right values for the strange
and bottom masses. Moreover we predict that the atmospheric mixing
angle is maximal, and $\theta_{13}\approx 0$ lepton mixing angle.

By using the value of the top, charm and up quark masses we predict
a small value for $\delta m^2_{sol}/\delta m^2_{atm}$ and for
the entries
$V_{ub}$, $V_{cb}$, $V_{td}$, and $V_{ts}$.
Due to a property coming from the $S_2$ structure of our model,
 they are all related to the small value of the ratio of the other quark
 masses with respect to $m_t$. 
On the other side when the $S_2$ symmetry is dynamically broken 
 the Cabibbo angle become relevant.

It is a pleasure for us to thank F.~Vissani for useful discussions
about limits and properties of $SO(10)$ models.
One of us (S.M.) would like to thank F.~Caravaglios for enlightening
discussion about permutation symmetries.

\end{document}